\documentclass[preprint]{elsarticle}
\usepackage[utf8]{inputenc}
\usepackage[a4paper, total={6.5in, 9.5in}]{geometry}
\usepackage{lmodern}
\usepackage{graphicx}
\usepackage[figurename=Fig.,labelfont=bf,labelsep=period]{caption}
\usepackage{subcaption}
\usepackage{amsmath}
\usepackage{amsfonts}
\usepackage{amssymb}
\usepackage{amsbsy}
\usepackage{newtxtext,newtxmath}
\usepackage[colorlinks=true,citecolor=red,linkcolor=black]{hyperref}
\usepackage{natbib}
\usepackage{multirow}

\usepackage{soul}
\usepackage[dvipsnames]{xcolor}
\usepackage{setspace}
\title{The Creation of Puffin, the Automatic Uncertainty Compiler}

\author[1,2]{Nicholas Gray\corref{cor1}}
\author[1]{Marco de Angelis}
\author[1]{Scott Ferson}
\cortext[cor1]{Corresponding author}
\address[1]{Institute for Risk and Uncertainty, University of Liverpool, Liverpool, United Kingdom, L69 7ZX}
\address[2]{nickgray@liverpool.ac.uk}
\begin{document}

\maketitle

\section*{Abstract}

An uncertainty compiler is a tool that automatically translates original computer source code lacking explicit uncertainty analysis into code containing appropriate uncertainty representations and uncertainty propagation algorithms. We have developed an prototype uncertainty compiler along with an associated object-oriented uncertainty language in the form of a stand-alone Python library. It handles the specifications of input uncertainties and inserts calls to intrusive uncertainty quantification algorithms in the library.
The uncertainty compiler can apply intrusive uncertainty propagation methods to codes or parts of codes and therefore more comprehensively and flexibly address both epistemic and aleatory uncertainties.

\textbf{Keywords}: Uncertainty Analysis; Uncertainty Compiler; Probability Bounds Analysis

\section{Introduction}

Modern science and engineering is all about numerical calculation. With the inexorable growth of computer power, more of these calculations are being undertaken with ever more complex computer simulations. These developments mean that new computation-intensive technologies are being explored, such as digital twins (see \cite[Sec.~2.2.3.3]{Shafto2012} or \cite{Boschert2016}). Scientists and engineers need to make calculations even when there is uncertainty about the quantities involved, yet the tools they are commonly using do not allow this to be done intrusively. As a result many analysts work with computer codes that do not take full account of uncertainties. 

Within the numerical calculations essential to engineering, there are two types of uncertainty: aleatory and epistemic. Aleatory uncertainty arises from the natural variability in changing environments and material properties, errors in manufacturing processes or inconsistencies in the realisations of systems. Aleatory uncertainty cannot be reduced by empirical effort. Epistemic uncertainty is caused by measurement imperfections or lack of perfect knowledge about a system. This could be due to not knowing the full specification of a system in the early phases of engineering design or ignorance about the expected manufacturing variations or deployment conditions. Imperfect scientific understanding of the underlying physics or biology involved causes uncertainty in predictions about the future performance of a system even after the design specifications have been decided. If uncertainties are small they can often be neglected or swept away by looking at the worst-case scenarios. However, in situations where the uncertainty is large, this approach is suboptimal or impossible, especially if it would impact a decision. Instead, a strategy of comprehensively accounting for the two kinds of uncertainty is needed that can propagate imprecise and variable numerical information through calculations. 

Because analysts are typically unwilling to rewrite their codes, various simple strategies have been used to remedy the problem, such as elaborate sensitivity studies or wrapping the program in a Monte Carlo loop. These approaches treat the program like a black box because users consider it uneditable. However, whenever it is possible to look inside the source code, it is better characterised as a crystal box because the operations involved are clear but fixed and unchangeable in the mind of the current user. 

\section{The Problem with Monte Carlo}

The most common approach to deal with uncertainty is to wrap code within a Monte Carlo shell. In this approach the calculations are repeated with random values for selected input variables.  This is done for a large number of iterations, and the distribution of resulting outputs can be analysed. Such tools exists in many programming langages: DAKOTA for C++ \cite{Adams2010}, COSSAN \cite{Patelli2015} and UQLab \cite{Marelli2014} for MATLAB or UQpy for Python \cite{Olivier2020}. Olivier et al. give an excellent overview of many more software packages that are availible for non-intrusive uncertainty quantification \cite{Olivier2020}. Under such an approach random values are chosen and then the calculations are performed and the output stored, this is done for a number of iterations and total outputs can be analysed after the process has been completed.

\begin{figure}[t]
    \centering
    \includegraphics[width=\textwidth]{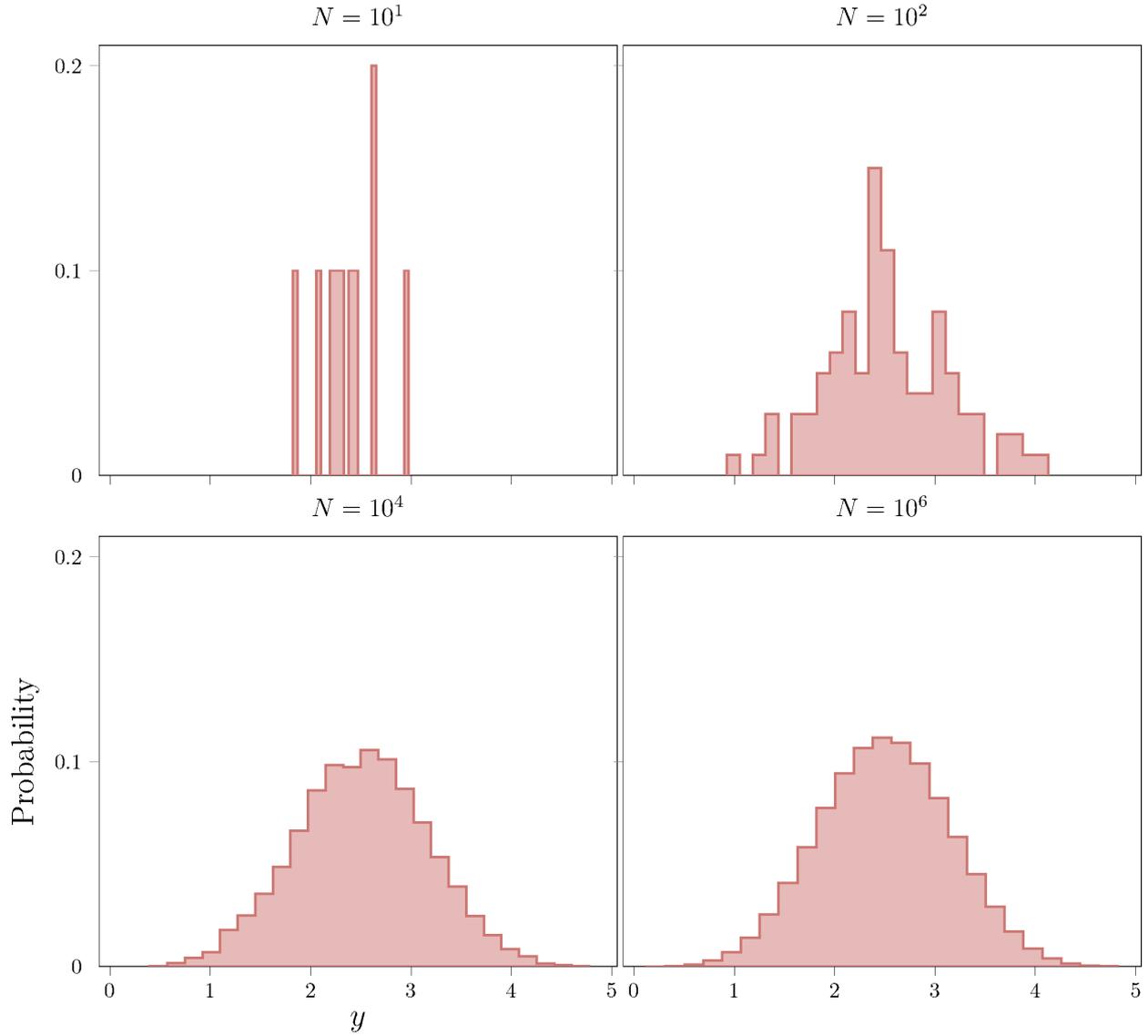}
    \caption{Normalised histogram for the Monte Carlo simulation of Equation~\ref{eq:MC} for increasing number of iterations.}
    \label{fig:MC1}
\end{figure}

In order to demonstrate the potential problems with such an approach we can consider a simple example. Suppose we have five variables $x_1,\dots,x_5$ which are known to all have a value between 0 and 1 but no further information is known about the values. Suppose we need to perform the calculation
\begin{equation}
    \label{eq:MC}
    y = x_1+x_2+x_3+x_4+x_5
\end{equation}
with the knowledge that some bad thing will happen if $y\ge4.5$. A number can be randomly generated for $x_1$, $x_2$, etc and these can be used in order to calculate the value of $y$ for $N$ iterations. After this is complete we can plot a histogram to show the distribution for $y$. Since we do not have any information about the distribution for $x_1,\dots,x_5$ it seems sensible to assume that all values are equally likely and use a uniform distribution. Figure~\ref{fig:MC1} shows these histograms for various $N$. From this we can see that as $N\rightarrow\infty$ the histogram resembles a normal distribution. 


Whatever the number of replications used in the Monte Carlo simulation, we can estimate that the probability of the bad thing happening. With $10^6$ replications, this estimate is $\Pr(y\ge4.5)=2.53\times 10^{-4}$. However, it seems reasonable to consider whether we have confidence that the event is so rare. We had no information about the distributions of the five values except that they were between 0 and 1.  Nor did we have knowledge about what dependencies there might be between the variables. From this information we cannot rule out the possibility that each $x$ value is much more likely to be closer to 1 than 0, or that there is some dependence between the $x$ values such that if $x_1$ is high then all the others are also likely to be high. Thus, the way that the uncertainty has been characterised may be significantly underestimating the risk \citep{Oberkampf2021}.

There have been several engineering failures that were due in part to underestimating risks in ways similar to this example \citep{Oberkampf2021,Pate-Cornell2012}. Before the 1986 Challenger Disaster, NASA management had predicted the probability of failure with loss of vehicle and crew as 1 in $10^5$ flights \cite{Feynman1986}. This turned out to be a gross underestimation of the true risk, which after the retirement of the fleet stood at 2 in 135. The Fukushima Daiichi nuclear disaster was due in part to underestimating the risk of a tsunami of the magnitude that caused the disaster
and in failing to understand that collocating the backup generators created dependence that destroyed the planned engineered redundancy
when the site was flooded during the event 
\citep[p.~48]{Amano2015}. The probabilities of satellites colliding in orbit can be underestimated through the use of probabilities \cite{Balch2019}, leading to false confidence that they are not going to hit each other.

Performing uncertainty analysis by simply wrapping a simulation code in a Monte Carlo loop may not give a full account of the uncertainties that are present within a simulation. The probabilities of extreme events
are especially difficult to correctly estimate when either the
distributions of input variables are not known or
any inter-variable dependencies are not known.
There are other limitations of this simplistic Monte Carlo 
approach, including false confidence \cite{Balch2019}, 
and problems arising from confounding epistemic and aleatory uncertainties \cite{Beer2013}.

\section{Puffin}

\begin{figure}
    \centering
    \includegraphics[width=0.5\textwidth]{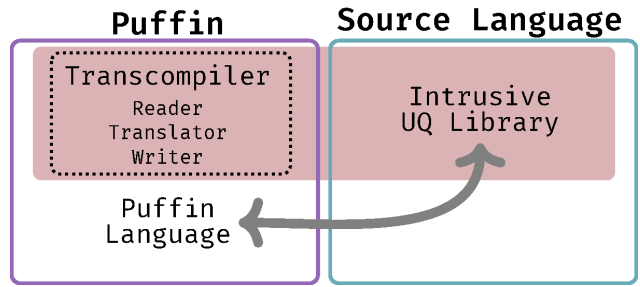}
    \caption{The different components of Puffin and the UQ library it depends on.  For each source language that Puffin is able to read, the parts highlighted in red need to be created. The intrusive uncertainty quantification library and Puffin Language need to mirror each other with a direct translation being available for all uncertainty specifications.}
    \label{fig:parts}
\end{figure}

Strategies are needed that automatically translate original source code into code with appropriate uncertainty representations and propagation algorithms. Perez et al. introduced a MATLAB toolbox to perform automatic uncertainty propagation using unscented transform, however more general approaches are needed \cite{Perez2020}. In this paper we describe an uncertainty compiler for this purpose, named Puffin, along with an associated language. It handles the specifications of input uncertainties and inserts calls to an object-oriented library of intrusive uncertainty quantification (UQ) algorithms. In theory, the approach could work with any computer language and any flavour of uncertainty propogation. There are several components that are needed for the creation of Puffin as shown in Figure~\ref{fig:parts}. 

Puffin needs a language of uncertainty, ``Puffin Language'', that allows users to specify what uncertainties should be associated with the variables within the source code. This language should be simple and independent of the source language. For every source language that Puffin is to work with, there has to be an intrusive UQ library that Puffin Language can be translated into. Puffin Language does not need to be a textual programming language, it could instead be a visual language as part of a graphical user interface. Every type of uncertainty that is expressible in Puffin Language must be supported by an object constructor in the uncertainty library written in the source language. Section~\ref{sec:puffin-language} discusses this Puffin Language component. 

The other component that Puffin needs is a transcompiler \cite{Ilyushin2016} that translates a user's source code into an UQ enriched code expressed in the same language. There are three subcomponents to this: a Reader that is able to read the source language, a Translator that is able to read the specified uncertainties in Puffin Language, and a Writer that combines the results of both to output a new script with the specified uncertainties. ANTLR, a parser/lexer generator, can be used to generate the Reader and Writer \citep{Parr2012}. ANTLR requires a grammar specification for the source language. Fortunately, ANTLR grammar files have been defined for many popular programming languages \cite{ANTLR-grammars} and these can be used as a starting point. The Reader scans the input script and identifies the assignment operators within that may have uncertainties that need to be specified. The Translator reads Puffin Language specifications created by the user and translates the uncertainties into the source language. For each source language these conversions need to be specified. The Writer reproduces the script with the required uncertainties and necessarily changes required for the analysis to run without issue.  We have designed these components for Python.

\section{Puffin Language}\label{sec:puffin-language}

Puffin depends on an uncertainty language to express what uncertainties are present within their scripts. This language enables users to specify the uncertainties about the variables involved in their code before compiling it into a new script with UQ enriched code. The language currently enables calculations with five types of uncertain objects that have relevance in engineering \citep{Beer2013}:

\begin{itemize}
\item Interval: unknown value or values for which sure bounds are known \citep{Moore2009},
\item Probability distribution: random values varying according to specified law such as normal, uniform, binomial, etc., with known parameters \citep{Ang2007}, 
\item Probability box: random values for which the probability distribution cannot be specified exactly but can be bounded \citep{Ferson2003},
\item Confidence structure: inferential uncertainty about a constant parameter compatible with both Bayesian and frequentest paradigms \citep{Balch2012}, and
\item Natural language expressions:  uncertain values indicated by linguistic hedge words such as `\texttt{about 7.2}' or `\texttt{at most 12}' \cite{Ferson2015,Jean2016,Lefort2017}.
\end{itemize}

Libraries that add some/all of these objects are available in C++ \citep{Mascarenhas2018}, Python \cite{pba.py}, MATLAB \cite{pba.matlab}, R [\citenum{pba.r} or \citenum{HYRISK}], Julia \cite{pba.jl}. There are many other uncertain objects that could be included within such a language of uncertainty such as
second-order distributions or meta-distributions \cite{Haenggi2021i,Haenggi2021ii}, fuzzy numbers \citep{Dijkman1983}, possibility distributions \citep{Dubois2005}, consonant structures \citep{Balch2020}, info-gap models \citep{Ben-Haim2006} and others.

\subsection{Intervals}
An interval is an uncertain number representing values from an unknown distribution over a specified range, or perhaps a single value that is imprecisely known even though it may in fact be fixed and unchanging. Intervals thus embody epistemic uncertainty. Intervals can be specified by a pair of scalars corresponding to the lower and upper bounds of the interval, such as $[0,1]$ or $[4,5]$. They can also be expressed as a value plus or minus some error, such as $[5 \pm 2]$ which is equivalent to $[3,7]$.

Interval arithmetic computes with ranges of possible values, as if many separate calculations were made under different scenarios. However, the actual computations made by the software are done all at once, so they are very efficient. Basic binary operations ($+,\ -,\ \times,\ \div$) can be performed using interval arithmetic:
\begin{equation}
  [a,b] + [c,d] = [a+c,b+d],
\end{equation}
\begin{equation}
  [a,b] - [c,d] = [a-d,b-c],
\end{equation}
\begin{equation}
  [a,b]\times[c,d] = \left[\mathrm{min}\left(ac,ad,bc,bd\right),\mathrm{max}\left(ac,ad,bc,bd\right)\right]
\end{equation}
and
\begin{equation}
  [a,b]\div[c,d] = \left[\mathrm{min}\left(\frac{a}{c},\frac{a}{d},\frac{b}{c},\frac{b}{d}\right),\mathrm{max}\left(\frac{a}{c},\frac{a}{d},\frac{b}{c},\frac{b}{d}\right) \right]
\end{equation}
assuming that $0 \not\in [c,d]$.

Intervals can be propagated through all common mathematical functions such as $\exp$, $\sin$, $\log$, etc. This is relatively straightforward if the function is monotonic as this implies that the endpoints of the input interval correspond to the endpoints of the output interval. For example, when calculating the exponential of an interval, 
\begin{equation}
    \exp{\left([0,1]\right)} = \left[\exp(0),\exp(1)\right] \approx \left[1,2.718\right]
\end{equation}
For a non-monotonic function such as sine it is not necessarily the case that the endpoints of the interval correspond the the endpoints of the output function. For example, it is not the case that
\begin{equation}
    \sin\left(\left[0,\pi\right]\right) = \left[\sin(0),\sin(\pi)\right] = [0,0]
\end{equation}
because there are many $x$ values such that $\sin(x)> 0$ for $x \in [0,\pi]$. The true width of the interval can be calculated using the maxima of the function within the domain, in this case $\frac{\pi}{2}$. Hence,
\begin{equation}
    \sin\left(\left[0,\pi\right]\right) = [0,1].
\end{equation}
\subsection{Probability Distributions and Probability Boxes}
A probability distribution is a mathematical function
that gives the probabilities of occurrence for different possible values of a random variable. 

\begin{figure}[t]
    \centering
    \includegraphics[width = 0.5\textwidth]{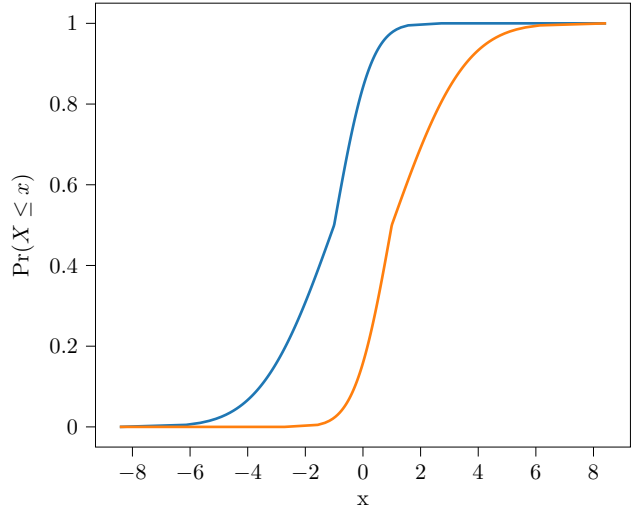}
    \caption{Probability box for a normal distribution with $\mu = [-1,1]$ and $\sigma=[1,2]$.}
    \label{fig:pbox}
\end{figure}

Probability bounds analysis integrates interval analysis and probability distributions using probability boxes (p-boxes) \citep{Ferson2003}. They can be considered as interval bounds on a probability distribution, therefore one can think of a probability distribution as a special case of a p-box. Figure~\ref{fig:pbox} shows an example of a probability distribution. P-boxes characterise both epistemic and aleatory uncertainty. A p-box can be expressed mathematically as
\begin{equation}
    \mathcal{F}(x) = [\underline{F}(x),\overline{F}(x)], \ \underline{F}(i)\geq \overline{F}(i)\ \forall x \in \mathbb{R}
\end{equation}
where $\underline{F}(x)$ is the function that defines the left bound of the p-box (the blue in Figure~\ref{fig:pbox}) and $\overline{F}(x)$ defines that right bound of the p-box (the orange line in Figure~\ref{fig:pbox})

As with intervals, standard arithmetic operations can be performed on p-boxes (and therefore probability distributions). For two p-boxes $\mathcal{A}(x) = [\underline{A}(x),\overline{A}(x)]$ and $\mathcal{B}(x) = [\underline{B}(x),\overline{B}(x)]$, then
\begin{equation} \label{eq:f1}
    \mathcal{C}(x) = \mathcal{A}(x) \circ \mathcal{B}(x) = [\underline{C(x)},\overline{C(x)}]
\end{equation}
where
\begin{subequations}\label{eq:f2}
\begin{align}
    \underline{C(z)} &= \inf_{z=x \circ y} \left[ \mathrm{min}\left( \underline{A(x)} \circ \underline{B(y)},1\right) \right] \\
    \overline{C(z)} &= \sup_{z=x \circ y} \left[ \mathrm{max}\left( \overline{A(x)} \circ \underline{B(y)}-1,0\right) \right]
\end{align}
\end{subequations}
if $\circ \in [+,\times]$, or
\begin{subequations}\label{eq:f3}
\begin{align}
    \underline{C(z)} &= 1+ \inf_{z=x \circ y} \left[ \mathrm{min}\left( \underline{A(x)} \circ \overline{B(y)} ,0\right) \right] \\
    \overline{C(z)} &= \sup_{z=x \circ y} \left[ \mathrm{max}\left( \overline{A(x)}  \circ \underline{B(y)},0\right) \right]
\end{align}
\end{subequations}
if $\circ \in [-,\div]$. Naturally, division is only valid if $0 \not \in \mathcal{B}$ \citep[p.~89]{Ferson2004}.

Within a programming language a p-box can be expressed by using the name of the probability distribution, or some shorthand for the name, as the function and the arguments as intervals. For example, the p-box shown in Figure~\ref{fig:pbox} could be generated using
\begin{center}
    \texttt{Normal([-1,1],[1,2])}.
\end{center}

P-boxes can be defined in situations where the shape of the distribution is unknown but some empirical evidence about the data is known, such as the minimum, maximum, mean, standard deviation, etc. In this situation bounds can be created such that they are consistent with all the available information \citep{Ferson2002}.

\subsection{Confidence Boxes}
Confidence boxes (c-boxes) are imprecise generalisations of traditional confidence distributions, which, like Student's $t$-distribution, encode frequentist confidence intervals for parameters of interest at every confidence level \citep{Balch2012,Ferson2013}. They are analogous to Bayesian posterior distributions in that they characterise the inferential uncertainty about distribution parameters estimated from sparse or imprecise sample data, but they have a purely frequentist interpretation that makes them useful in engineering because they offer a guarantee of statistical performance through repeated use. Unlike confidence intervals which cannot usually be used in mathematical calculations, c-boxes can be propagated through mathematical expressions using the ordinary machinery of probability bounds analysis, and this allows analysts to compute with confidence, both figuratively and literally, because the results also have the same confidence interpretation \cite{Wimbush2021}. For instance, they can be used to compute probability boxes for both prediction and tolerance distributions. Confidence boxes can be computed in a variety of ways directly from random sample data. There are c-boxes both for parametric problems (where the family of the underlying distribution from which the data was randomly generated is known to be normal, binomial, Poisson, etc.), and for non-parametric problems in which the shape of the underlying distribution is unknown. C-boxes account for the uncertainty about a parameter that comes from the inference about observations, including the effect of small sample size, but also the effects of imprecision in the data and demographic uncertainty which arises from trying to characterise a continuous parameter from discrete data observations.

For example, it is possible to specify a c-box in the binomial case of having $K$ successes out of $N$ trials, based upon Clopper-Pearson confidence intervals \citep{Balch2020,Brown2001,Clopper1934}. This $K$-out-of-$N$ c-box is specified as
\begin{equation}
  \label{eq:KN}
  \mathcal{KN}(k,n) = \left[\mathrm{beta}(k,n-k+1),\mathrm{beta}(k+1,n-k)\right].
\end{equation}

\subsection{Natural Language Uncertainty}
In order to make uncertainty analysis as simple as possible, users should be able to input their uncertainties using natural language expressions such as \textit{about} or \textit{almost}. Humans are more likely to express their uncertainties in terms of hedged expressions around a round number, rather than as a percentage or probability. Table~\ref{tab:hedge} lists some hedge words and their possible interpretations. Hedge words can be interpreted as intervals, p-boxes \citep{Ferson2015}, or consonant c-boxes \citep{Hose2021}.

\begin{table}[h]
    \begin{center}
        \begin{tabular}{|c|c|}
            \hline
            Hedged Numerical Expression & Possible Interpretation \\ \hline
            $\textsc{about}(x)$ & $ \lbrack x \pm 2 \times 10^{-d} \rbrack $ \\
            $\textsc{around}(x)$ & $ \lbrack x \pm 10 \times 10^{-d} \rbrack $ \\
            $\textsc{count}(x)$ & $ \lbrack x \pm \sqrt{x} \rbrack $ \\
            $\textsc{almost}(x)$ & $ \lbrack x - 0.5 \times 10^{-d},x \rbrack $ \\
            $\textsc{over}(x)$ & $ \lbrack x, x + 0.5 \times 10^{-d} \rbrack $ \\
            $\textsc{above}(x)$ & $ \lbrack x, x + 2 \times 10^{-d} \rbrack $ \\
            $\textsc{below}(x)$ & $ \lbrack x - 2 \times 10^{-d}, x \rbrack $ \\
            $\textsc{at most}(x)$ & $ \lbrack 0,x \rbrack $ \\
            $\textsc{at least}(x)$ & $ \lbrack x,\infty \rbrack$ \\
            $\textsc{order}(x)$ & $ \lbrack x/2, 5x \rbrack $ \\
            $\textsc{between}\ x\ \textsc{and}\ y$ & $ \lbrack x,y \rbrack $ \\
            $K\ \textsc{out of}\ N$ & $\left[\mathrm{beta}(k+1,n-k),\mathrm{beta}(k,n-k+1) \right]$\\ \hline
        \end{tabular}
    \end{center}
    \caption{Hedge expressions and their mathematical equivalent. Note: $d$ is the number of significant figures of $x$.}
    \label{tab:hedge}
\end{table}

\subsection{Logical Operations with Uncertain Objects} \label{sec:logic}
When making decisions it is often the case that two values need to be compared with each other. Asking whether an observed value is greater than, equal to, or less than some threshold value is fundamental. For example, if a decision relies on some observed value $X$ being less than 1, when we know the value of $X$ accurately then it is easy to make such a comparison. However, if there is some uncertainty about the value of $X$ then this comparison may not be so easy.

For intervals, $X = [a,b]$ and $Y=[c,d]$, then
\begin{equation}
    X < Y = \begin{cases}
        1\ &b < c \\
        0\ &a \geq c\\
        [0,1]\ &\mathrm{otherwise}
    \end{cases}
\end{equation}
and 
\begin{equation}
    X > Y = \begin{cases}
        0\ &b \leq c \\
        1\ &a > d\\
        [0,1]\ &\mathrm{otherwise}
    \end{cases}
\end{equation}
with 0 and 1 denoting true and false respectively, and [0,1] being the Boolean equivalent of ``I don't know''. We can call [0,1] the \textit{dunno} interval. This implies that we cannot say whether an uncertain value characterised by an interval is larger or smaller than another unless the interval is entirely greater or less than the other interval. 
For the equality comparison,
\begin{equation}
    X == Y = \begin{cases}
        [0,1]\ & a \in Y \ \mathrm{or} \ b \in Y\\
        0\ &\mathrm{otherwise},
    \end{cases}
\end{equation}
when asking for equivalence between intervals it is never possible to say that one value is equal to another. We can introduce a new Boolean operator (===) to test for whether two uncertain numbers are equivalent in form,
\begin{equation}
    X === Y = \begin{cases}
        1\ & a = c \ \mathrm{and}\ b=d\\
        0\ &\mathrm{otherwise.}
    \end{cases}
\end{equation}
The dunno interval can be converted into a true Boolean using operators such as \textsc{always} or \textsc{sometimes}
\begin{subequations}
    \begin{align}
        \textsc{always}\left([0,1]\right) &= 0 \\
        \textsc{sometimes}\left([0,1]\right) &= 1
    \end{align}
\end{subequations}
so that we can get
\begin{equation}
\label{eq:always}
    \textsc{always}\left(X<Y\right) = \begin{cases}
        1\ &b < c \\
        0\ &\mathrm{otherwise}
    \end{cases}
\end{equation}
\begin{equation}
\label{eq:sometimes}
    \textsc{sometimes}\left(X<Y\right) = \begin{cases}
        1\ &a < d \\
        0\ &\mathrm{otherwise}.
    \end{cases}
\end{equation}
There are methods that are able to deal with more nuanced ways of using logical operations with intervals, see \cite{Kreinovich2014} as an example. There are also different logic systems such as fuzzy logic that could be used in order to make logical operations with uncertain numbers \citep{Zadeh1988}.

\subsection{Repeated Variables and Dependency} \label{sec:dependence}
When performing intrusive uncertainty analysis it would be ideal to always obtain best possible results that are guaranteed to bound the true value without overestimating the uncertainty. The uncertainty can be inflated or artifactually high if careful consideration of the dependence between, and repetition of, uncertain numbers is not undertaken. This problem appears to be ubiquitous to many, if not all, uncertainty calculi \citep{Beer2013}. 

For example, if $a = [2,3]$ and $b = [4,5]$, then $a \times b = [8,15]$. However, if it were the case that $a$ and $b$ were oppositely dependent on each other, such that a low value of $a$ is always matched with a high value of $b$, then $a \times b$ is the much narrower interval $[10,12]$.

Repetition of variables can also artifactually inflate the amount of uncertainty present within the output. For example if $a = [1,2]$, $b = [-1,1]$ and $c = [3,4]$ then
\begin{equation} \label{eq:abac}
    \begin{split}
    ab+ac &= [1,2]\times[-1,1] + [1,2]\times[3,4]\\
          &= [-2,2] + [3,8]\\
          &= [1,10]
    \end{split}
\end{equation}
but
\begin{equation} \label{eq:abc}
    \begin{split}
  a(b+c) &= [1,2]\left([-1,1]+[3,4]\right) \\
         &= [1,2]\times[2,5] \\
         &= [2,10]
    \end{split}
\end{equation}
Although algebraically these two expressions should be equal,  the uncertainty of $ab+ac$ is greater than the uncertainty about $a(b+c)$. This is due to the fact that the uncertain variable $a$ is repeated within the former but appears only once in the latter. In essence the uncertainty about $a$ has been considered twice when performing the first calculation. The amount of this artifactual uncertainty can be reduced by transforming the original equation into a single-use expression where uncertain variables are only used once. If this is not possible, there are other techniques that can be used to reduce this artifactual uncertainty (e.g. \cite{DeFigueiredo2004,Goubault2016,Ander2021,Kramer2006})

\begin{figure}[t]
    \centering
    \includegraphics[width = 0.75\textwidth]{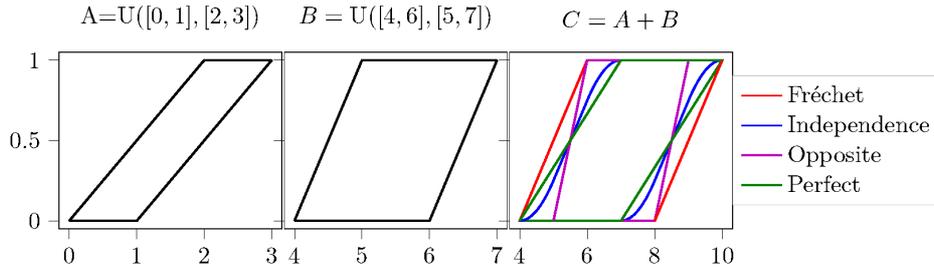}
    \caption{Addition of two p-boxes with different dependencies.}
    \label{fig:pbox_dep_add}
\end{figure}

For distributions and p-boxes, significant artifactual uncertainty reduction can be made if the dependence between the variables is known \citep{Ferson2004}. Figure~\ref{fig:pbox_dep_add} shows the result of adding two separate p-boxes, $A=\mathrm{U}([0,1],[2,3])$, $B=\mathrm{U}([4,6],[5,7])$, together with different dependencies between $A$ and $B$. The Fr\'echet bounds are used when the dependence between $A$ and $B$ is unknown, thus it is the most general case and is guaranteed to bound the correct answer. As such in Figure~\ref{fig:pbox_dep_add} the Fr\'echet bounds cover all the other dependencies, as it is the operation that is defined in equations \ref{eq:f1}, \ref{eq:f2} and \ref{eq:f3}. Perfect, or comonotonic, is where there is perfect positive dependence between the two variables, with the highest possible correlation coefficient. Opposite, or countermonotonic, is perfect negative dependence between the two variable with the lowest possible correlation coefficient. Independence is where there is no dependence between the two variables. It should not be assumed that variables are independent unless this is known because wrongly assuming independence can lead to incorrectly reducing the amount of uncertainty and understating tail risks. 

In general, dependence between uncertain quantities can be expressed through the use of correlation coefficients or copulas or bounds on copulas more generally \cite{Embrechts2003,Nelsen2006,Joe2014,Ander2021-Copulas}. This can include named copulas such as independence, opposite and perfect, as shown in Figure~\ref{fig:pbox_dep_add}, or other copula families parameterised by a numerical correlation coefficient. Independence implies the correlation is zero, although zero correlation does not imply independence.  Likewise, a correlation of one implies perfect dependence, but, depending on the copula family, perfect dependence may not imply correlation one. The symbol $\equiv$ can be used to indicate that the variables are equal in value, i.e., equal in distribution and perfectly positively correlated. 

These dependencies can be stored within a matrix, such as that shown in Table~\ref{tab:dependency_table}. This matrix can be checked for feasibility by checking that it is positive semi-definite and that there are no conflicting dependencies within the table. For example the matrix shown in Table~\ref{tab:bad_dependencies} is not logically consistent for continuous variables. This is because a high value of \texttt{x} implies a high value of \texttt{z}, since they are positively dependent on each other. Meanwhile, a high \texttt{z} implies \texttt{y} must be low, due to their opposite dependence. This means there must also be dependence between \texttt{x} and \texttt{y}, not the independence as has been specified within the table.

\begin{table}[h!]
    \centering
    \begin{tabular}{|c|ccccc|}
        \hline     & \texttt{a} & \texttt{b} & \texttt{c} & \texttt{d} & \texttt{e} \\ \hline
        \texttt{a} &            & i          & $\equiv$   & f          & o          \\
        \texttt{b} & i          &            & i          & f          & 0          \\
        \texttt{c} & $\equiv$   & i          &            & f          & o          \\
        \texttt{d} & f          & f          & f          &            & f          \\ 
        \texttt{e} & o          & 0          & o          & f          &            \\ \hline
    \end{tabular}
    \caption{Matrix showing dependencies between several variables. (f - Fr\'echet, i - Independence, o - Opposite, $\equiv$ - Equal in value)}
    \label{tab:dependency_table}
\end{table}
\begin{table}[h!]
    \centering
    \begin{tabular}{|c|ccc|}
        \hline     & \texttt{x} & \texttt{y} & \texttt{z}  \\ \hline
        \texttt{x} &            & i          & p           \\
        \texttt{y} & i          &            & o           \\
        \texttt{z} & p          & o          &             \\ \hline
    \end{tabular}
    \caption{Dependency matrix that does not make logical sense. (i - Independence, p - Perfect, o - Opposite)}
    \label{tab:bad_dependencies}
\end{table}

\subsection{Other issues}
Aside from dependencies and sensitivity to repeated variables, there are other issues that distinguish simple deterministic calculations from uncertainty quantifications.
For example, in uncertainty quantification analysts may need to consider \emph{ensembles} and \emph{backcalculations}.

Probability distributions describe properties or behaviours across a population of entities. Statisticians call such a population the “reference class” or “ensemble”. Uncertainty quantification implicitly represents many calculations over interacting ensembles, and it can be extremely important to keep in mind what the values in a probability distribution represent.  For instance, if the post-operative risks of prostatectomy is fifty percent erectile dysfunction, it would make a huge difference to a patient whether this means that 50\% of his future attempts at sex fail or that 50\% of patients are permanently impotent.  Does a system reach 10\% of criticality or does it reach criticality 10\% of the time? Uncertainty quantifications that do not explicitly define what the distributions in an analysis represent in terms of their respective ensembles may be meaningless.
Puffin allows users to annotate their codes to specify and document the ensemble described by any distribution or other uncertain quantity, although, in general, it is the responsibility of the analyst to ensure that the calculations used make sense.

Another wrinkle that makes uncertainty quantification different from its analogous deterministic calculations is the importance of backcalculation.
Backcalculation is a mathematical operation for finding solutions to equations involving variability or uncertainty that guarantee some desired performance. Such problems are ubiquitous in engineering design.  Backcalculation solves questions such as 
\begin{enumerate}
    \item What dimensional constraints on a component are necessary to ensure that it fits in its place in a machine given spatial tolerances?
    \item How much propellant is needed to guarantee sufficient fuel given the mission contingencies and unforeseen variabilities?
    \item How much shielding is needed on a spacecraft to ensure that the total ionizing radiation experienced inside the craft does not exceed some tolerable threshold, given that radiation in space varies over time in an imperfectly known way?
\end{enumerate}  
The Puffin UQ library
%
%
%
%
%
%
%
%
has algorithms to solve backcalculations that involve intervals, distributions and p-boxes (when solutions exist), but it is the responsibility of the analyst to ensure that they are deployed 
appropriately to yield calculations that make sense in the engineering context.

\section{Compiler}
Puffin consists of its intrusive UQ library, a code inspector/editor, and an uncertainty compiler.  
Puffin's uncertainty compiler does five things:
\begin{enumerate}
    \item Parses the input source code into expression tree(s),
    \item Identifies the variables in any assignment operations,
    \item Replaces or modifies some or all of these assignments in the expression trees according to options and specifications provided by the user,
    \item Translates the expression trees, with amended assignments, into the target language equipped with its intrusive UQ library, and
    \item Analyses the output code to detect repeated variables and other functional dependencies that affect calculations and suggests improvements for computing uncertainties. 
\end{enumerate}

\begin{figure}[t]
    \centering
    \includegraphics[width = 0.75\textwidth]{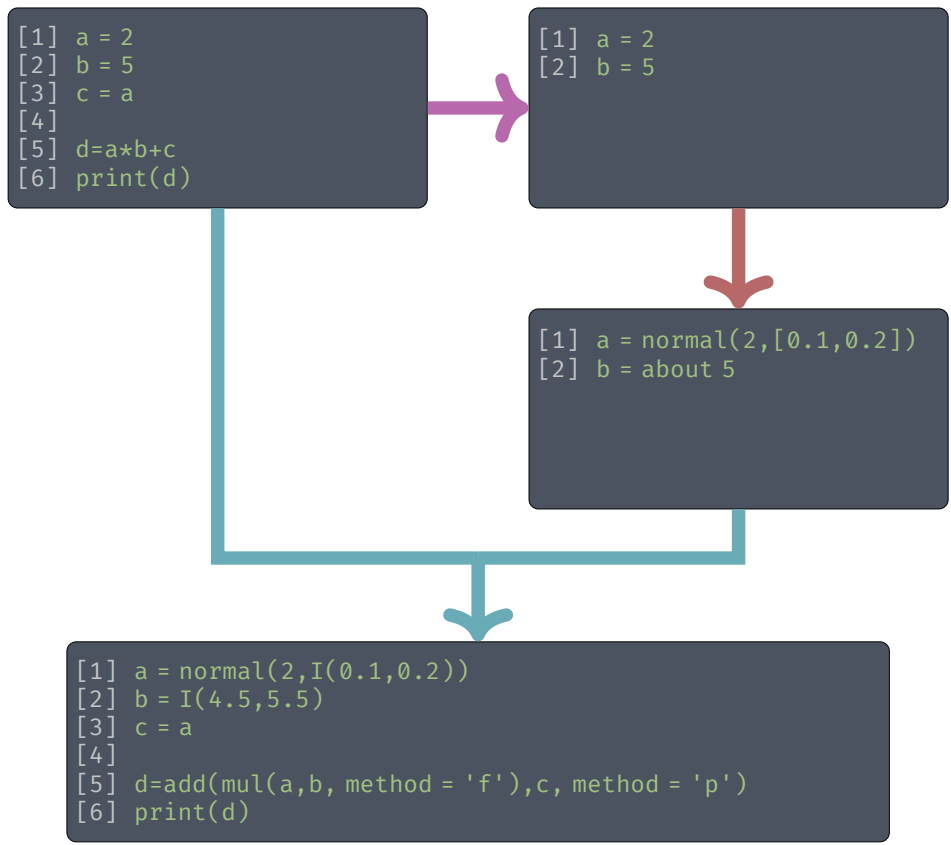}
    \caption{The result of using the compiler whilst defining the uncertainty in Puffin on a simple pseudocode script.}
    \label{fig:Puffin_w_uq}
\end{figure}

\begin{figure}[t]
    \centering
    \includegraphics[width = 0.75\textwidth]{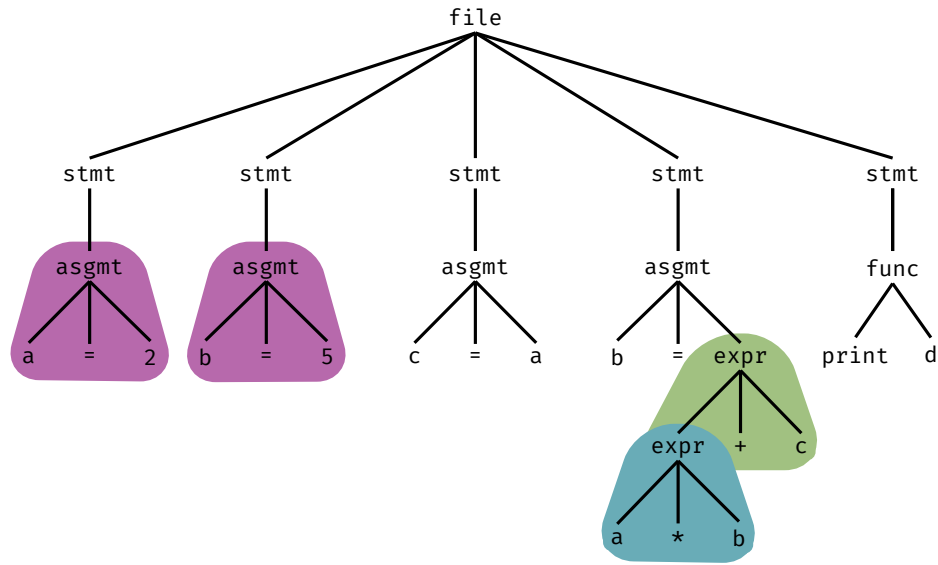}
    \caption{Parse tree for the simple pseudocode script. Abbreviations: stmt - statement, asgmt - assignment, func - function, expr - expression.}
    \label{fig:parse-tree}
\end{figure}

In order to explain what is happening in these steps it is useful to consider a simple pseudocode script, shown in the top left corner of Figure~\ref{fig:Puffin_w_uq}. For step 1, the simple script has then been broken into a parse tree which can be seen in Figure~\ref{fig:parse-tree}. From this tree Puffin detects the assignment operators which define a variable. 
These include lines 1 and 2, the leaves highlighted magenta on the parse tree, but not those that assign a value based upon a mathematical expression (line 5), a function or directly from another variable (line 3). 
In theory, such variables could also be edited by the user, but Puffin assumes that only explicit assignments will have uncertainty.
Once the assignments have been found they can be displayed in the Puffin language, shown in the top right panel of Figure~\ref{fig:Puffin_w_uq} where the assignments can then be edited with the appropriate user-specified uncertainties. These uncertainties need to then be translated to the source language, along with the rest of the parse tree. 

This translation may include altering any functions that depend the amended variables.
In this case, the infix operators in the definition of \texttt{d} in line 5 (\texttt{+,\textasteriskcentered}) have been identified within the parse tree and replaced with an explicit call to the UQ library functions (\texttt{add}, \texttt{mul}) which also have as an argument the dependence operation that is to be used. The lower panel of 
Figure~\ref{fig:Puffin_w_uq}, the value 'f' of this argument corresponds to making no assumption about the intervariable dependence between \texttt{a} and \texttt{b}, and perfect dependence (comonotonity) between their product and the variable \texttt{c}.

Puffin should only highlight numeric objects, not characters, strings or other non-numeric classes. 
In strongly typed programming languages like C, FORTRAN, and Pascal, the problem of distinguishing numeric from other types of objects is easy.  In Python, R or Julia, the type of any object is not detectable until runtime and can even change during execution.
Puffin will also need to be able to recognise objects that are collections of numeric values such as lists or dictionaries. It will also need to be able to look at what is inside the lists and highlight those which have numeric objects within the list and allow of the individual objects to have uncertainties added. Alternatively it could be the case that the whole of the list has the same uncertainty, something which should be possible.
%
%
%
%

\begin{figure}[t]
    \centering
    \includegraphics[width = 0.75\textwidth]{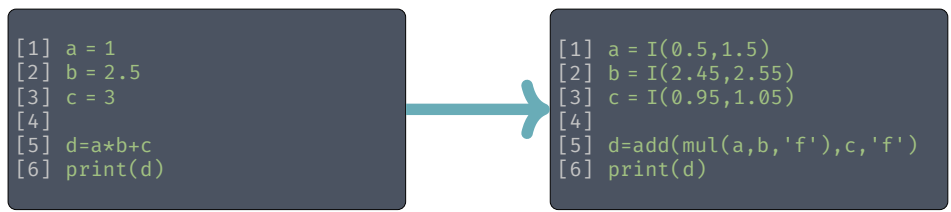}
    \caption{Result of using Puffin automatically on a simple pseudocode script.}
    \label{fig:Puffin_auto}
\end{figure}

Puffin can be run automatically without any user input at all. Under default settings, automatic uncertainty compilation replaces floating-point constants with intervals interpreted from the significant figures used in the source code assignments and uses that information as a proxy for the uncertainty (for an example see Figure~\ref{fig:Puffin_auto}). In this mode all the steps of the steps happen concurrently without requiring any further input from an end user. When using this mode the compiler will need to tread carefully around mathematical constants such as $\pi$ or $e$ for which there is no uncertainty. Ideally it would allow users to minimally specify what values are precise constants. 

\subsection{Control Flow and Functions}

\begin{figure}[t]
    \centering
    \includegraphics[width = 0.75\textwidth]{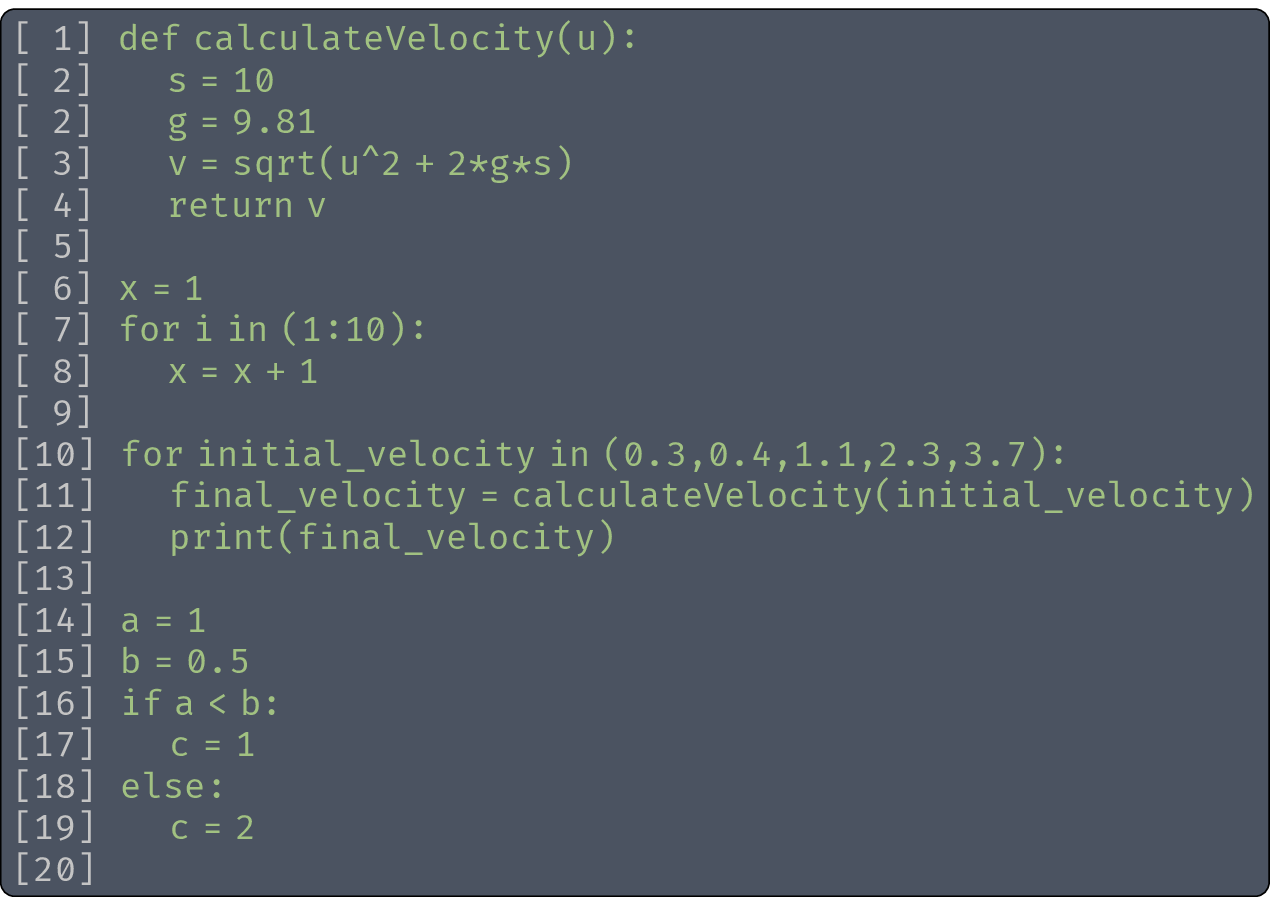}
    \caption{Pseudocode script with functions and for loops.}
    \label{fig:cff}
\end{figure}

For loops and functions here are potential stumbling blocks for Puffin. Figure~\ref{fig:cff} shows a simple pseudocode script with a function and a for loop. The first for loop each \texttt{i} is simply just a control variable with start and end variable. The individual value of \texttt{i} is irrelevant and as such would have no uncertainty about it. The second for loop is a `for each' loop implying that the code needs to do something for each value within some iterable object, under this scenario it may be the case that there is uncertainty about the object within the list. In this case the code is setting the value of \texttt{initial\_velocity} as each value within the the list, for each iteration of the for loop. In this case it may be the case that there is uncertainty within the object, in which case Puffin should recognise this and all users to change the code such that the objects within the list can have uncertainty added to them. 

Puffin will also need to have a way of dealing with local variables within functions. For example, the function in Figure~\ref{fig:cff} has two local variables, \texttt{s} for the distance that the object travels and \texttt{g} for the acceleration due to gravity. It is conceivable that both of these variables have some uncertainty associated with them and as such Puffin should be able to detect the variables and offer the ability to edit them so that uncertainty is handled. This could be done using a dot notation, meaning that the local \texttt{g} can be accessed using \texttt{calculateVelocity.g}. 

If statements and other logical control structures may also pose issues for Puffin. In line 16 of Figure~\ref{fig:cff} the logical operation within the statement would need changing to ensure that the statement runs as expected, see Section~\ref{sec:logic}. Ideally, the analyst would decide what should happen if the statement \texttt{a < b} returned a dunno, [0,1], result by using the adverb operators discussed in equations \ref{eq:always} and \ref{eq:sometimes}. This may require additional editing to deal with situations where an uncertain result should be handled differently to a certain true or false. 

All the code and variables are not necessarily contained within one script.  For example, classes and functions are often placed in other files in order to improve readability or to avoid repetitions. Ideally Puffin would be able to parse several scripts at the same time and remember the context for all the individual objects. It is also often the case that scripts read data from other files when running. Under this scenario it may be difficult to use Puffin to express the uncertainty directly within the script, although import functions could be modified to add in the uncertainties. 
For instance, anytime a floating-point number is read from the file, its significant digits could be interpreted to specify an interval around the value.  So, for example, the value `3.56' would be understood as the interval [3.555, 3.565].
Another approach might be to get Puffin to parse the data file and add the uncertainties in to the file directly. This would require changing the import function to be able to handle uncertain datafiles. 

Many computers languages that are not purely functional support functions that specify their parameters with "call by reference" which means that the memory location of a value is passed to the function rather than a copy of the true value. This convention can allow the function to change the values of those parameters in the calling routine not just locally within the function. Python does this by default with objects more complicated than integers, floating-point number, and strings such as lists, dataframes and numpy arrays. Puffin will need to be careful in handling functions that use the call by reference method of passing argument.
% ...

The presence of uncertainty implies multiple function definitions might be useful. For instance \texttt{sqrt} applied to ranges that might include negative numbers could have three possible behaviours. Abnormal termination, for example Python's \texttt{math.sqrt} returns a domain error if passed a negative number; yielding imaginary results, such as Python's \texttt{cmath.sqrt}; or ignoring the negative values, returning [0,1] for $\sqrt{[-1,1]}$\footnote{In this instance the square brackets correspond to Python's list object not an interval}. An example of this can be seen when using \texttt{numpy.sqrt([-1,1])} in Python which returns [NaN, 1]. Similar facilities already exist to handle NaN's or missing values. 

%

\subsection{Coping with Dependency and Repeated Variable Problems} \label{sec:rvp2}
\begin{figure}[t]
    \centering
    \includegraphics{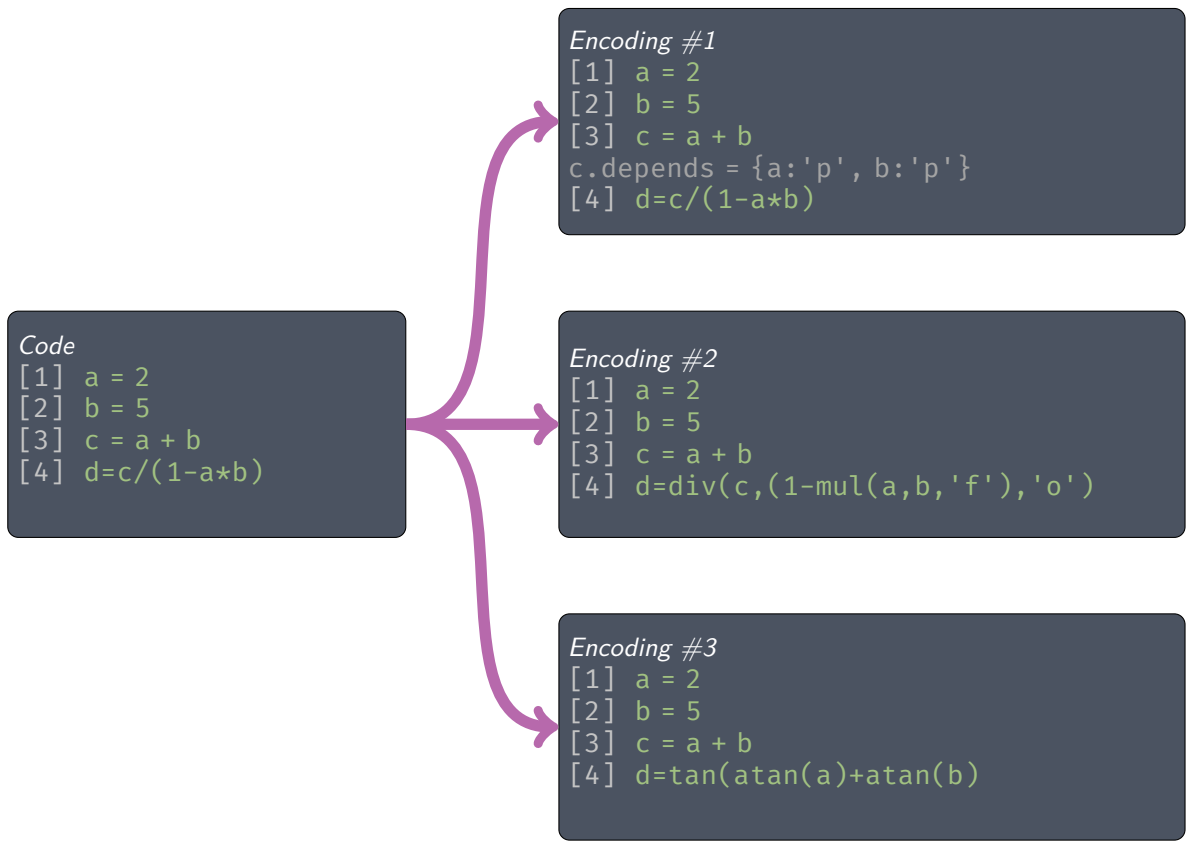}
    \caption{Three possible encodings for the dependencies in the code shown on the left.}
    \label{fig:puffin_dependence}
\end{figure}

When it comes to dealing with the issues of dependency and repeated variables there are a couple of approaches that could be used in order to help reduced the problems discussed in Section~\ref{sec:dependence}. The simplest approach from a Puffin perspective would be for the libraries within each language to be able to handle the dependencies directly. This could be done if each object kept track of what other objects it depends on and in what way. For the example that has been used in Figure~\ref{fig:puffin_dependence}--Encoding \#1, the variable \texttt{c} would remember that it is dependent on the variables \texttt{a} and \texttt{b} and therefore on line 4 it would know what the correct arithmetic would be to ensure as little artifactual uncertainty as possible. Puffin would insert this dependence directly in the translation as shown with the grey text in the Encoding \#1, then at run time the / and \textasteriskcentered operators would automatically invoke the correct algorithms that respect the dependencies between the variables detected. This approach would have demands on memory. It requires initialised variables to have dependencies specified or a default dependence if they are unspecified. 

The other way of treating the dependence would be for Puffin to parse over the script in order to detect the dependencies directly at compile time. These dependencies can then be stored within a matrix as discussed in Section~\ref{sec:dependence}. This matrix would need to be accessible for an analyst to add in assumptions about dependencies not observable from the code. For example, if variables \texttt{a} and \texttt{b} are independent this cannot be directly inferred from the code and therefore Fr\'echet would be assumed unless the analyst stated otherwise by editing the matrix. In this scenario Puffin would have to replace any infix operators with function calls with the appropriate dependence. In Figure~\ref{fig:puffin_dependence}--Encoding \#2, the multiplication \texttt{a}\textasteriskcentered\texttt{b} has been replaced by the multiply function specifying that Fr\'echet should be used for the dependencies. Similarly the division operator has been replaced by a function with the method defined as opposite.

Finally where it is possible Puffin should be able to rearrange the equations such that any repeated variables are removed. Such a method would require Puffin to have a directory of multi-use to single-use rearrangements as well as a way of matching the written code to the mathematical expression. Another, smarter, approach would be to have a symbolic algebra system that is able to rearrange to a single-use expression on the fly. The simplest version of this is for it to happen just across one line, for instance replacing 
\begin{center}
    \texttt{c = a\textasteriskcentered b+a\textasteriskcentered c}
\end{center}
with 
\begin{center}
    \texttt{c = a\textasteriskcentered(b+c)}.
\end{center}
A more complex approach is to consider repetitions globally to try to reduce the repetitions by detecting repetitions that happen over multiple lines. For instance in the example code in Figure~\ref{fig:puffin_dependence}, there is a hidden repetition across lines 3 and 4 because $c=a+b$. So
\begin{equation}
    d = \frac{c}{1-a*b} = \frac{a+b}{1-a*b}.
\end{equation}
This expression can be rearranged into the single-use expression 
\begin{equation}
    d = \tan\left( \arctan(a) + \arctan(b) \right)
\end{equation}
which is the change made in Encoding \#3. Such a transformation would be in the directory mentioned above.

Care would need to be taken to ensure that the right variable gets the rearrangement. Take the following kinematics equation to find the position $s$ of a particle at time $t$
\begin{equation}
    s = ut + \frac{1}{2}at^2
    \label{eq:suvat}
\end{equation}
where  $u$ is the initial velocity and $a$ is the acceleration of the particle. This equation has a single repetition for both $u$ and $a$ but $t$ is repeated. If there is uncertainty associated with $t$ then this equation can be rearranged into a single-use expression
\begin{equation}
    s = \left(\sqrt{\frac{a}{2}}t+\frac{u}{\sqrt{2a}}\  \right)^2 - \frac{u^2}{2a}.
    \label{eq:suvat_sue}
\end{equation}
This equation contains repetitions of $a$ and $u$ and as such may only be preferred if there is no uncertainty associated with either $a$ or $u$. If there is uncertainty associated with either then it may be best not to perform the rearrangement or to intersect possible rearrangements to obtain the best possible expression.

There are additional issues that Puffin could face when rearranging equations. For example if the uncertainty about $a$ includes negative numbers then $\sqrt{2a}$ is likely to be problematic. Alternatively, there could be problems if $a$ straddles 0 because this would result in a division by zero. A strategy for dealing with this may be to perform the calculations in using both Equation~\ref{eq:suvat} and \ref{eq:suvat_sue} and intersect them.

\subsection{Hermeneutic Problems} \label{sec:hermenutical}
There are several problems that could occur when it comes to translating a script because it is difficult to understand the intent of the programmer from the code.

An example of this can be found in line 3 of the psuedocode example in Figure~\ref{fig:Puffin_w_uq}. There is potential for confusion when it comes to the assignment \texttt{c = a} as there are a couple of different interpretations as to what such a command implies when it comes to the uncertainty. The first is that we are implying that \texttt{c} and \texttt{a} are the same object but have been given different names for some reason, under this scenario they should be considered equivalent to each other and therefore the calculation \texttt{a + c} could be rearranged to the single use expression \texttt{2\textasteriskcentered a}. A second interpretation is to consider that the line could have been written as \texttt{c = 1\textasteriskcentered a} and the \texttt{1} has been dropped as it would have had no mathematical impact on the calculation, this implies that they are perfectly dependent on each other in the same way that \texttt{c = -1\textasteriskcentered a} implies negative dependence. The calculation \texttt{a + c} would therefore need to be performed using perfect dependence. A third interpretation would be to consider that it is saying that \texttt{c} is a copy of \texttt{a}, they have the same uncertainty but their realisations are not necessarily related to each other but they have the same distribution shape. In the third scenario it would be sensible to make no assumptions about the dependencies between \texttt{a} and \texttt{c} and therefore Fr\'echet should be used. Knowledge of which of these scenarios is correct depends on the context of the script, something which Puffin is unable to make an assumption about by itself.

\begin{figure}[t]
    \centering
    \begin{subfigure}[]{0.45\textwidth}
        \includegraphics[width = 0.95\textwidth]{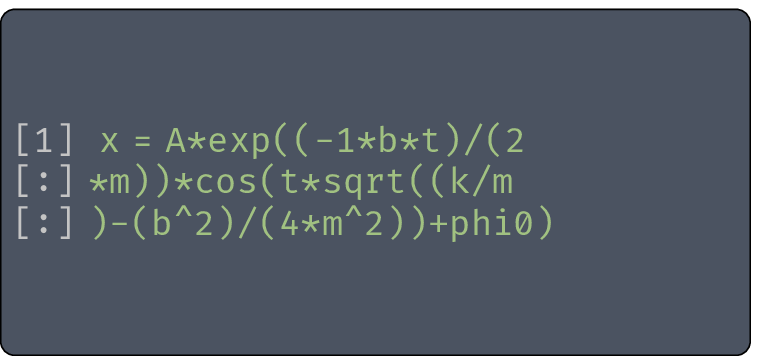}      
        \caption{Coded in a single line}
        \label{fig:readability1}
    \end{subfigure}
    \begin{subfigure}[]{0.45\textwidth}
        \includegraphics[width = 0.95\textwidth]{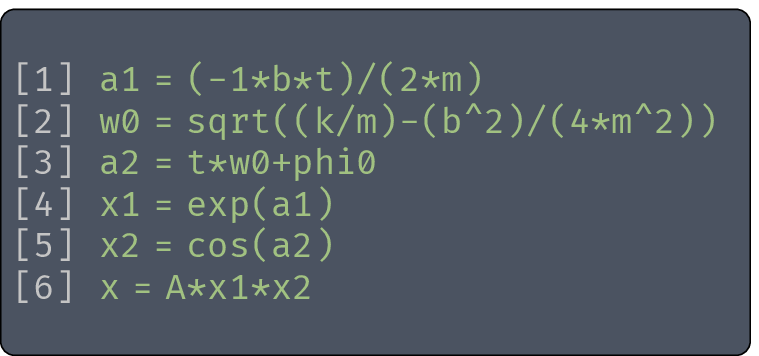}      
        \caption{Coded across multiple lines}
        \label{fig:readability2}
    \end{subfigure}
    \caption{\label{fig:readability} Two different way in which Equation~\ref{eq:DHO} might be coded.}
\end{figure}

Another potential interpretation problem can occur because when creating code people naturally favour making their code readable. For example, the equation of motion for a damped harmonic oscillator can be given by 
\begin{equation}
    x''(t) + \frac{b}{m} x'(t) + \frac{k}{m} x(t) = 0
    \label{eq:DHOdiff}
\end{equation}
where $b$ is a damping constant, $m$ is the mass of the oscillator and $k$ is the spring constant. This equation can be solved analytically to find
\begin{equation}
    x(t) = A\exp{\left(\frac{-bt}{2m}\right)}\cos{\left(t\sqrt{\frac{k}{m}-\frac{b^2}{4m^2}}+\phi_0\right)}
    \label{eq:DHO}
\end{equation}
where $A$ is a constant and $\phi_0$ is the initial angle. In Figure~\ref{fig:readability} the equation has been coded in two different ways. In \ref{fig:readability1} the equation has been coded on a single line, as the equation is quite complicated it is likely that the programmer who is coding the equation would want to split it into multiple parts as has been done in \ref{fig:readability2}. There are no mathematical difference between the two approaches as they will lead to the same value. Puffin however would have to be careful around breaking up equations in such a way, strategies would be needed to ensure that breaking the lines up would not have a detrimental effect on how the code operated. Care would also need to be taken about the dependency tracking throughout the split equation, from example if there was uncertainty about the damping constant $b$ then it would be difficult to assess the dependence between lines 4 and 5, especially since the cosine function is not monotonic. It may be better to use other techniques to solve the ODE (Equation~\ref{eq:DHOdiff}) such as VSOPE or VNODE \citep{Nedialkov1999,Nedialkov2001,Nedialkov2006,Lin2007,Enszer2015}. Making such a change would again require knowledge of what the calculation is and what exactly it is doing, something unlikely to be obvious from parsing the code.

\section{Discussion}
There are several reasons why Puffin may be unlikely to work as intended. The view that deterministic calculations can be translated to analogous computations for uncertainty quantification simply by replacing point values with uncertain structures like intervals and distributions ignores the issues of
\begin{enumerate}
    \item computational burden,
    \item repeated variables,
    \item dependencies,
    \item input specification, 
    \item conditionals, 
    \item ensembles, and 
    \item backcalculations. 
\end{enumerate}
This paper has described several strategies to mitigate the complexities arising from these issues. 
It is also likely to be the case that Puffin would not be able to introduce the most perfect uncertainty translations. Manually editing code or creating a new uncertainty aware script from scatch will likely outperform the automatic changes made by Puffin. This problem would not be unique to Puffin however, in general hand coding is always likely to outperform source-to-source translation \cite{Plaisted2013}. 

Puffin is almost the exact opposite of an optimising compiler, which aims to minimise a programme's execution time and memory requirements. By replacing objects with uncertain equivalents both of these will almost certainly increase. For instance, if intervalising calculations increases the computational time 5- to 20-fold, clearly a simulation limited by computational time would need to be scaled back.  Distributions or p-boxes would be still more burdensome.  Of course, efficiency is not always a critical issue and this extra computational effort does pay for global uncertainty propagation and what computer scientists call automatic result verification \citep{Adams1993}.  Moreover, Puffin's implementation of modern uncertainty quantification could be more efficient and more comprehensive than simply embedding a deterministic computation inside a Monte Carlo shell with millions of replications.

It is unrealistic to expect that simply replacing  computations involving integer and floating-point variables with analogous imprecise computations involving corresponding intervals, distributions and p-boxes will yield correct and useful results. 
The repeated variable problem and, more generally, the dependency problem, discussed in sections~\ref{sec:dependence}~and~\ref{sec:rvp2}, can artifactually inflate the uncertainty of a complex computation.  Even if the calculations are technically correct in the sense that they enclose the true uncertainty, the naive application of intrusive uncertainty quantification can sometimes yield results with massively inflated uncertainty that renders them practically useless.  Having massive uncertainty is not the problem itself; the true uncertainty may actually be large.  The problem is when it the uncertainty artifactually depends on the way the analysis was structured and does not reflect the features of the underlying computational problem.

The analysts who most need Puffin are those who have never heard of a p-box and who aren't sure what normal distributions or intervals are.  Puffin offers multiple ways of specifying uncertainties in order to cater to the needs of such analysts. Puffin offers multiple ways of specifying uncertain inputs for user unfamiliar with the details that they require:
\begin{itemize}
    \item Significant-digit intervals 
    \item Measurement intervals (manufacturer or GUM conventions)
    \item English-language hedge words (`about', `less than', etc.)
    \item Poisson model counts
    \item Moment or moment-range specifications
    \item Equivalent binomial count (k out of n confidence box)
    \item Single-sample confidence intervals
    \item Mean normal range (n \& range normal distribution)
    \item Fermi strategies
    \item Distribution-free specification of p-boxes
\end{itemize}
It is also possible to run Puffin in a way that doesn't require any inputs, transforming assignments into significant digit intervals.  Irrespective of how the uncertainties are expressed within Puffin, it is worth remembering that uncertain garbage in will lead to uncertain garbage out.


The general problem of uncertainty analysis is hard and it is difficult to create software that comprehensively solves all these problems and the development if Puffin is likely to be difficult. However, many practical problems are simpler than the most general problem and when this is the case it would be extremely useful to use a tool like Puffin which is able to handle uncertainty analysis intrusively within code. Puffin is intended to be open source  and to be continuously co-developed by an interested community, adding in functionality and extending it as fashionable programming languages change and uncertainty quantification techniques develop.

\section*{Code Availability}
Puffin is currently in development, and the current version can be found on GitHub\cite{Puffin-github}. We welcome suggestions and collaborators via GitHub.

\section*{Acknowledgements}
This work has been funded by the Engineering and Physical Science Research Council (EPSRC) through the programme grant ``Digital twins for improved dynamic design'', EP/R006768/1.

\bibliographystyle{elsarticle-num}

\end{document}